\documentstyle[14pt]{article}

\begin{document}           
\title{Report IRB-StP-GR-260596 \newline \newline \newline
{\bf
An Example of Exact Solution in the LTB Model
}
}
\author{\it by \\
\\
{\bf Alexander Gromov}
\\ \\
\small\it St. Petersburg State Technical University \\
\small\it Faculty of Technical Cybernetics, Dept. of Computer Science \\
\small\it 29, Polytechnicheskaya str. St.-Petersburg, 195251, Russia \\
\small and \\
\small\it Istituto per la Ricerca di Base \\
\small\it Castello Principe Pignatelli del Comune di Monteroduni \\
\small\it I-86075 Monteroduni(IS), Molise, Italia \\
\small\it e-mail: gromov@natus.stud.pu.ru
}
\date{}
\maketitle
\begin{abstract}
The Caushy problem in the LTB model is formulated.
The rules of calculating three undetermined functions which defined a
solution in the LTB model are presented.
One example of exact nonhomogeneous
model is studied. The limit transformation to the
FRW model is shown.
\\ \\
PACS number(s):98.80
\end{abstract}
\newpage
\section{The Introduction} \label{introd}

The LTB model is one of the most known spherical symmetry model in general
relativity. It was created by
Lemaitre \cite{Lemaitre:33}, Tolman \cite{Tolman:34}, and Bondy
\cite{Bondy:47}
during the period of time from 1933 to 1947. The exact
solution have been obtained by Bonnor \cite{Bonnor:72} in 1972 and
\cite{Bonnor:74} in 1974.
The LTB model represented one of the simplest nonhomogeneous nonstationary
cosmological models and due to this fact
is used to study some new ideas in the cosmology.

The present interest to LTB was risen by the observations
shown the fractal structure of the Universe in the
large scale \cite{Coleman},
\cite{Pietronero:87}. The review of the topics in Cosmology is presented
in \cite{Bar}.
In a set of papers the LTB model is used to study the
observational datas and
redshift as a main cosmological test \cite{Ribeiro:92a} -
\cite{Moffat:94b}.

The central problem of
using the LTB model is in calculation of three undetermined
functions which defined the solution. There is a set of ways how to solve
this problem from the physical point of view in the mentioned papers.
This article is devoted to the mathematical point of view on this matter.

\section{The LTB Model} \label{T_sol}

This section is devoted to presentation the Lemaitre-Tolman-Bondy model
and section {2} of the paper \cite{Tolman:34} is cited.
The co-moving system of coordinate is used in this model where the interval
has the form
\begin{equation}
ds^2(r,t) = -e^{\lambda(r,t)}dr^2 - e^{\omega(r,t)}
\left(
d\theta^2 + sin^2\theta d\phi
\right) + dt^2.
\label{ds2}
\end{equation}
$\lambda(r,t)$ and $\omega(r,t)$ are metrical functions \cite{Tolman:34}
definding the solution.
In the co-moving system of coordinate with the line element (\ref{ds2})
the energy-momentum tensor
\begin{equation}
T^{\alpha,\beta} = \rho \frac{dx^{\alpha}}{ds}\frac{dx^{\beta}}{ds}
\label{T_ik}
\end{equation}
has only one non zero component
\begin{displaymath}
T^{4}_{4} = \rho \qquad
T^{\alpha}_{\beta} = 0, \qquad \alpha \quad \mbox{or} \quad \beta =
4.
\label{T_ik:RO}
\end{displaymath}
Using them together with Dingle results \cite{Dingle} we obtain the system
of equations of the LTB model:
\begin{equation}
8 \pi T^{1}_{1} = e^{-\omega} -
e^{-\lambda}\frac{{\omega^{\prime}}^2}{4} + \ddot \omega +
\frac{3}{4} \dot \omega^2 - \Lambda = 0
\label{T:4}
\end{equation}
\begin{eqnarray}
8 \pi T^{2}_{2} = 8 \pi T^{3}_{3} = \nonumber\\
- e^{- \lambda}
\left(
\frac{\omega^{\prime\prime}}{2} +
\frac{\omega^{\prime}{}^{2}}{4} -
\frac{\lambda^{\prime}\omega^{\prime}}{}
\right)
+\frac{\ddot \lambda}{4}
+\frac{\dot\lambda^2}{4}
+ \frac{\ddot \omega}{2}
+\frac{\dot \omega^2}{4}
+\frac{\dot \lambda \dot \omega}{4}
- \Lambda = 0
\label{T:5}
\end{eqnarray}
\begin{equation}
8 \pi T^{4}_{4} =e^{-\omega} - e^{-\lambda}
\left(
\ddot \omega +\frac{3}{4}\dot \omega^2 - \frac{\lambda^{\prime}
\omega^{\prime}}{2}
\right)
+ \frac{\dot \omega^2}{2}
+ \frac{\dot \lambda \dot \omega}{2}
- \Lambda = 8 \pi \rho
\label{T:6}
\end{equation}
\begin{equation}
8 \pi e^{\lambda} T^{1}_{4} = - 8 \pi T^{4}_{1} =
\frac{\omega^{\prime}\dot \omega}{2}
- \frac{\dot \lambda \omega^{\prime}}{2}
+\dot \omega^{\prime} = 0,
\label{T:7}
\end{equation}
where
\begin{equation}
{}^{\prime} = \frac{\partial}{\partial r} \qquad
\dot{} = \frac{\partial}{\partial t}
\label{T:dot_prime}
\end{equation}
The equation (\ref{T:7}) has the solution
\begin{equation}
e^{\lambda} = e^{\omega} \frac{\omega^{\prime \,2}}{4 f^2(r)}
,
\label{T:8}
\end{equation}
where $f(r)$ is undetermined function.
Substituting (\ref{T:8}) into (\ref{T:4}) we obtain
\begin{equation}
e^{\omega}
\left(
\ddot \omega +\frac{3}{4} \dot \omega^2 - \Lambda
\right)
+
\left[
1 - f^2(r)
\right] = 0.
\label{T:10}
\end{equation}
This equation is integrated twice.
First integral gives the equation
\begin{equation}
e^{3 \,\omega / 2}
\left(
\frac{\dot \omega^2}{2} - \frac{2}{3} \Lambda
\right)
+2 e^{\omega /2} \left[1 - f^2(r)\right] = F(r),
\label{T:11}
\end{equation}
and the second one gives the equation
\begin{equation}
\int
\frac
{{\rm d e^{\omega / 2}}}
{
\sqrt{{\rm
f^2(r) - 1 +\frac{1}{2} } F {\rm (r) e^{-\omega/2} + \frac{\Lambda}{3}
e^{\omega}}
}
}
 = t +  {\bf F} {\rm (r)
}
\label{T:12}
\end{equation}
The equations (\ref{T:11}) and (\ref{T:12}) hold undetermined functions
$\it F(r)$ и $\bf F\it(r)$.
The substitution of (\ref{T:8}) into (\ref{T:6}) together with
(\ref{T:11}) gives the equation for density
\begin{equation}
8 \pi \rho = \frac{1}{\omega^{\prime}e^{3\omega/2}}
\frac{
\partial
F{\rm (r)}}{{\rm \partial r}}
\label{T:15}
\end{equation}
%


\section{The Cauchy Problem for the LTB Model}

Before we study the LTB model let us introduce the follow
characteris\-tic values:
a velocity of light $c$, an observational meaning of the Habble
constant $\tilde H$, a characteristic time $1/\tilde H$ and characteristic
length $c/ \tilde H$.
We use the co-moving system of coordinates in the LTB model, so the radial
coordinate $r$ has a sense of Lagrangian mass coordinate
\cite{L&L},\cite{LE&AD}.
Two dimensionless variables $\mu$ and $\tau$ are definded by the rules
\begin{equation}
\mu = \frac{r}{m} \quad \quad \quad
\tau = \tilde H t,
\label{new_var}
\end{equation}
where $m$ is a full mass of the "gas".
The dimensionless Habble function $h(\mu,\tau)$ and density
$\delta(\mu,\tau)$ will be also used:
\begin{equation}
h(\mu,\tau) = \frac{H(\mu,\tau)}{\tilde H}, \qquad
\delta(\mu,\tau) = \frac{\rho(r,t)}{\rho_0},
\label{h}
\end{equation}
where $\rho_0 = \rho(0,0)$, $H(0,0) = \tilde{H}$.

Let us write the interval (\ref{ds2}) as
\begin{equation}
ds^2(r,t) = -A e^{\lambda(r,t)}dr^2 - B e^{\omega(r,t)}
\left(
d\theta^2 + sin^2\theta d\phi
\right) +c^2dt^2,
\label{ds2_1}
\end{equation}
where two constants $A$ and $B$
are introduced to take into account
the fact that (\ref{ds2_1}) is dimension equation.

The dimension of $\left[ ds^2 \right]$ is $L^2$, dimension of
$\left[ A \right]$ is $L^2 M^{-2}$ and dimension of $\left[ B \right]$ is
$L^2$, so
\begin{equation}
A = \left(\frac{c}{\tilde H m}\right)^2, \qquad B =
\left(\frac{c}{\tilde H}\right)^2.
\label{ds2_coeff}
\end{equation}

The interval (\ref{ds2_1}) has now the form
\begin{equation}
\left(\frac{\tilde H}{c}\right)^2ds^2(r,t) =
-e^{\lambda(r,t)}dr^2 - e^{\omega(r,t)}
\left(
d\theta^2 + sin^2\theta d\phi
\right) +d\tau^2,
\label{ds2_D-less}
\end{equation}
Together with metrical functions $\omega(\mu,\tau)$ and $\lambda(\mu,\tau)$,
introduced by Tolman, it is conveniently to use
the Bonnor's function \cite{Bonnor:72}
\begin{equation}
R(\mu,\tau) = e^{\omega(\mu,\tau)/2}.
\label{Bonnor}
\end{equation}
In the Bonnor's notation the interval (\ref{ds2_D-less}) takes the form
\begin{equation}
\left(\frac{\tilde H}{c}\right)^2 ds^2(\mu,\tau) =
-\frac{\left[R^{\prime}(\mu,\tau)\right]^2}{f^2} d\mu^2 -
R^2(\mu,\tau)
\left(
d\theta^2 + sin^2\theta d\phi
\right) + d\tau^2
\label{ds2_new}
\end{equation}
As it is shown in \cite{L&L} and \cite{LE&AD}, the Bonnor's coordinate
$R(\mu,\tau)$ has
a sense of Euler coordinate, so the equation (\ref{Bonnor})
correlates geometrical radius of the sphere $R(\mu,\tau)$ where the
particle is located, and the Lagrangian coordinate $\mu$ of this sphere.

To describe the radial motion we will use the Habble function connected
with variation of the length $d l$:
\begin{equation}
h = \frac{d \dot l}{d l},
\label{habble_def}
\end{equation}
where, according to the (\ref{ds2_D-less}), for $dl^2$ we read:
\begin{equation}
dl^2 = e^{\lambda(\mu,\tau)} d \mu^2  =
\left(
\frac{R^{\prime}}{f} d\mu
\right)^2.
\label{d l}
\end{equation}
By the substitution (\ref{d l}) into the definition of the Habble
function (\ref{Habble_def}) we obtain
\begin{equation}
h(\mu,\tau) = \frac{\dot \lambda(\mu,\tau)}{2} = \frac{\dot R
^{\prime}}{R^{\prime}}
= \frac{\partial \ln{R^{\prime}}}{\partial \tau}
\label{Habble_def}
\end{equation}
By the integration of the equation (\ref{Habble_def})
we obtain the formula for metrical function $\lambda(\mu,\tau)$:
\begin{equation}
\lambda(\mu,\tau) = 2 \int\limits_{0}^{\tau}h(\mu,\tau)d\tau +
\lambda(\mu,0)
\label{lambda_def_Habbl}
\end{equation}

A solution in the LTB model is defined by the
functions $f(r)$, $\it F(r)$ and $\bf F\it(r)$.
These functions are obtained in the process of solution of the system of
PDE, so to define them the initial/boundary conditions should
definitely be used.
The metrical function $\omega(\mu,\tau)$ is the solution of the equation
(\ref{T:12}).
The equations of the LTB model are obtained in \cite{Tolman:34}
and solved in the parametric form
for the three cases $f^2(\mu) < 1$, $f^2(\mu) = 1$ and $f^2(\mu) >
1$ in \cite{Bonnor:72} and \cite{Bonnor:74}.

The equations (\ref{T:10}) - (\ref{T:12}) are valid for every $\tau$,
and due to this fact in the Cauchy problem they {\it definde} the functions
$f^2(\mu)$, $\it F(\mu)$ and $\bf F\it(\mu)$ at the moment of time
$\tau = 0$:
\begin{equation}
{\rm
f^2(\mu) - 1 =
e^{\omega_0(\mu)}
\left(
\ddot \omega_0(\mu) + \frac{3}{4} \dot \omega^2_0(\mu) - \Lambda
\right)
}
\label{DEF:f^2-1}
\end{equation}
\begin{equation}
F{\rm(\mu)} =
{\rm
e^{3 \,\omega_0(\mu) / 2}
\left(
\frac{\dot \omega^2_0(\mu)}{2} - \frac{2}{3} \Lambda
\right)
+2
e^{\omega_0(\mu) /2}
\left[1 - f^2(\mu)\right]}
\label{DEF:itF}
\end{equation}
\begin{equation}
{\bf F} {\rm (\mu)} =
{\rm
\int \limits_{
e^{\omega_0(0)/2}
}^{
e^{\omega_0(\mu)/2}
} }
\frac{
d e^{\tilde\omega_0(\mu) / 2}
      }
     {
      \sqrt{
f^2(\mu) - 1 +\frac{1}{2} F (\mu) e^{-\tilde\omega_0(\mu)/2} +
            \frac{\Lambda}{3}
e^{\tilde\omega_0(\mu)}
           }
     }
\label{DEF:F}
\end{equation}
At the time $\tau = 0$ the equation (\ref{T:8}) {\it defindes} the function
$\lambda_0(\mu)$:
\begin{equation}
e^{\lambda_0(\mu)} =
e^{\omega_0(\mu)}
\frac{[\omega_0(\mu)]^{\prime \,2}}{4f^2(\mu)}.
\label{DEF:lambda_0_new)}
\end{equation}
Substituting (\ref{DEF:f^2-1}) into (\ref{DEF:itF}), we obtain
\begin{equation}
F
{\rm
(\mu) = e^{3\omega_0(\mu)/2}\left(
-2 \ddot\omega_0(\mu) - \dot\omega^2_0(\mu) +
\frac{4}{3}\Lambda
\right)
}.
\label{DEF:itF_new}
\end{equation}

Comparing (\ref{T:8}) - (\ref{T:11}) with (\ref{DEF:f^2-1}) -
(\ref{DEF:itF}), we find out that
%
\begin{eqnarray}
f^2(\mu) - 1 =
e^{\omega_0(\mu)}
\left(
\ddot \omega_0(\mu) + \frac{3}{4} \dot \omega^2_0(\mu) - \Lambda
\right) &=& \nonumber\\
e^{\omega(\mu,\tau)}
\left(
\ddot \omega(\mu,\tau) + \frac{3}{4} \dot \omega^2(\mu,\tau) - \Lambda
\right)
\label{DEF:f^2-1_n}
\end{eqnarray}
and
\begin{eqnarray}
F(\mu) =
e^{3\omega_0(\mu)/2}\left(
-2 \ddot\omega_0(\mu) - \dot\omega^2_0(\mu) +
\frac{4}{3}\Lambda
\right) &=& \nonumber\\
e^{3\omega(\mu,\tau)/2}\left(
-2 \ddot\omega(\mu,\tau) - \dot\omega^2(\mu,\tau) +
\frac{4}{3}\Lambda
\right)
\label{DEF:itF_new_n}
\end{eqnarray}
are not dependent on time.
Let's use the previous results to calculate the functions ${\bf F}(\mu)$
and integral in the equation (\ref{T:12}).
Substituting the definitions (\ref{DEF:f^2-1}) and (\ref{DEF:itF_new}) into
(\ref{DEF:F}) we obtain:
\begin{equation}
{\bf F}(\mu) =
\pm \int \limits_{\omega_0(0)}^{\omega_0(\mu)}
\frac{d \tilde\omega}{\dot{\tilde\omega}}.
\label{DEF:F_new}
\end{equation}

The function ${\bf F}(\mu)$ is equal to zero at the moment of time
$\tau = 0$ according the definition.
Substituting the right part of the equations (\ref{DEF:f^2-1_n}) and
(\ref{DEF:itF_new_n}) into the (\ref{DEF:F}), we obtain the equation
\begin{equation}
\pm\int \limits^{\omega(\mu,\tau)}_{\omega(\mu,0)}
\frac{d \tilde\omega}{\dot{\tilde\omega}} =
\pm \int \limits_{\omega_0(0)}^{\omega_0(\mu)}
\frac{d \tilde\omega}{\dot{\tilde\omega}} + \tau.
\label{DEF:F_new_nn}
\end{equation}

This analysis of the LTB model shows that the functions
\begin{equation}
\left.
\begin{array}{c}
%
%
\left.\omega(\mu,\tau)\right|_{\tau=0} = \omega_0(\mu) \quad
%
%
\left.\dot\omega(\mu,\tau)\right|_{\tau=0} = \dot\omega_0(\mu) \quad \\ \\
\left.\ddot\omega(\mu,\tau)\right|_{\tau=0} = \ddot\omega_0(\mu),
\end{array}
\right\}
\label{init}
\end{equation}
and constants
\begin{equation}
\left.
\begin{array}{c}
%
%
\left.\omega(\mu,0)\right|_{\mu=0} = \omega_0(0) \quad
%
%
\left.\dot\omega(\mu,0)\right|_{\mu=0} = \dot\omega_0(0) \quad \\ \\
\left.\ddot\omega(\mu,0)\right|_{\mu=0} = \ddot\omega_0(0), \quad
\Lambda,
\end{array}
\right\}
\label{init}
\end{equation}
are included into the definitions (\ref{DEF:f^2-1}) - (\ref{DEF:F})
and they form the initial conditions of the Cauchy problem for the
equations (\ref{T:4}) - (\ref{T:7}).
In accordance with (\ref{DEF:lambda_0_new)})
the function $\lambda_0(\mu)$ is not include in
the set of initial conditions.

Substituting (\ref{DEF:itF_new}) into the (\ref{T:15}),
we obtain the general expression for the density of "gas" in the LTB model:
\begin{eqnarray}
{\rm
    8 \pi \rho(\mu,\tau) =
    \frac{e^{
             \frac{3}{2}
             [\omega_0(\mu) - \omega(\mu,\tau)]
            }
         }
         {
          \omega^{\prime}(\mu,\tau)
         }
} \times
\nonumber\\
{\rm
    \left\{
           3\left[\omega_0(\mu)\right]^{\prime}
            \left[
                  -\ddot\omega_0(\mu) - \frac{1}{2}\dot\omega_0^2(\mu) +
                  \frac{\Lambda}{6}
            \right]
     - 2\left[\ddot\omega_0(\mu)\right]^{\prime} -
     2\dot\omega_0(\mu)\left[\dot\omega_0(\mu)\right]^{\prime}
    \right\}
}
\label{DEF:rho_new}
\end{eqnarray}
The function $\omega(\mu,\tau)$ from the equation (\ref{DEF:rho_new})
is the solution of the equation \cite{Tolman:34}.


\section{The $f^2 = 1$, $\Lambda = 0$ solution}

The solution of the main equation of the LTB model (\ref{T:12})
will be obtained in this section for one special case
\begin{equation}
f^2 = 1, \qquad \Lambda = 0.
\label{cond}
\end{equation}
This case allows simple analytical solution.
The law of the density, Habble function, and cosmological parameter  will
be also studied here.

The condition
$f^2(\mu) = 1$, as it goes from the (\ref{init}),
correlates the initial conditions as follows:
\begin{equation}
\ddot \omega_0(\mu) + \frac{3}{4} \dot \omega^2_0(\mu) = 0.
\label{init_f}
\end{equation}
Due to this reason,
the general number of the initial condition (\ref{init}) is decreased by
one unit.
To obtain the solution we'll start with the solution of the equation
(\ref{T:11}).
>From it it's obvious that:
\begin{equation}
\left(
\frac{\partial e^{\omega/2}}{\partial \tau}
\right)^2 =
f^2 - 1 + \frac{1}{2} F e^{-\omega/2} + \frac{\Lambda}{3} e^{\omega}.
\label{dif_1}
\end{equation}
In the general case the function $F$ is defined by the equation
(\ref{DEF:itF}).
Whit the help of (\ref{cond}) it takes the form
\begin{equation}
F = \frac{\dot\omega_0^2}{2}e^{3\omega_0/2}.
\label{F_1}
\end{equation}
The equation (\ref{dif_1}) with conditions (\ref{cond}) takes the form
\begin{equation}
e^{3\omega_0/4}\frac{
\partial e^{3\omega/4}}{\partial \tau}
= \pm\frac{3}{4}\dot\omega_0.
\label{solution_1}
\end{equation}
Integrating (\ref{solution_1}) we obtain
\begin{equation}
e^{\frac{3}{4}[\omega - \omega_0]} = \pm\frac{3}{4}\dot\omega_0\tau + {\bf
F}(\mu).
\label{solution_2}
\end{equation}
We find out the function {\bf F} from the initial conditions
then time $\tau = 0$:
\begin{equation}
{\bf F} = 1.
\label{F F}
\end{equation}
The solution now is:

\begin{equation}
e^{\frac{3}{4}[\omega - \omega_0]} = 1 \pm\frac{3}{4}\dot\omega_0\tau.
\label{solution_3}
\end{equation}
In the Bonnor's notation the equation (\ref{DEF:F_f_1})
becomes as follows:
\begin{equation}
\frac{R(\mu,\tau)}{R_0(\mu)} = \left(
1 \pm\frac{3}{4}\dot\omega_0(\mu)\tau
\right)^{2/3},
\label{DEF:F_f_1}
\end{equation}
where the upper sign correspond to the expansion into the infinity
from the initial condition
and lower sign correspond to the collapse from one.
The solution with "-" describes time of collapse depening on the initial
mass coordinate: if the particle has a Euler coordinate $R_0(\mu)$ at the
moment $\tau = 0$, the time of collapse is equal to
\begin{equation}
\bar\tau(\mu,0) = \frac{4}{3\dot\omega_0(\mu)}
\label{time}
\end{equation}
We obtain the density corresponding to this solution by the substitution
(\ref{solution_3}) into (\ref{DEF:rho_new}). Let's denote
\begin{equation}
\nu = \frac{[\dot\omega_0]^{\prime} }{ {\omega_0}^{\prime} }.
\label{nu-def}
\end{equation}
\begin{equation}
8 \pi \delta(\mu,\tau) =
\displaystyle
\frac{\dot\omega_0}{\left(
1 \pm\frac{3}{4}\dot\omega_0\tau
\right)^2} \ \cdot
\displaystyle
\frac{
\nu + \frac{3}{4} \dot\omega_0
}
{
1 \pm
\displaystyle
\frac{\nu\tau}{
1 \pm\frac{3}{4}\dot\omega_0\tau
}
},
\label{ro-sol}
\end{equation}

Using the definition (\ref{Habble_def}) we find out the Habble's function
for the solution (\ref{solution_3}):
\begin{equation}
h(\mu,\tau) = \pm\frac{\frac{3}{4}\dot\omega_0 + \nu}
            {1 \pm \left(\frac{3}{4}\dot\omega_0 +\nu\right)\tau}
\mp
\frac{1}{4} \, \cdot
\frac{\dot\omega_0}
     {1 \pm \frac{3}{4}\dot\omega_0\tau}
\label{Habbl-sol}
\end{equation}
Knowing the law of the density and Habble's function we obtain now
the formulum for the cosmological parameter. By the definition
\begin{equation}
\Omega = \frac{\rho}{\rho_c}, \quad \mbox {where} \quad
\rho_c = \frac{3H_0^2}{8\pi G},
\label{omega}
\end{equation}
where $H_0$ and $\rho$ mean the Habble function and the density at the
moment of the observation.
Let's assume that this moment is $\tau = 0$.
The density, the critical density and Habble function at this moment are:
\begin{equation}
8 \pi \delta(\mu,0) =
\dot\omega_0 \left(\frac{3}{4}\dot\omega_0 + \nu\right),
\label{ro-sol-a}
\end{equation}
\begin{equation}
\delta_{c}(\mu,0) = \frac{3 \tilde{H}^2}{8 \pi G \rho_0} h^2(\mu,0).
\label{ro-cr-dl}
\end{equation}
\begin{equation}
h(\mu,0) = \pm\left(\frac{1}{2}\dot\omega_0 +\nu\right)
\label{Habbl-sol-12}
\end{equation}
\begin{equation}
\Omega(\mu,0) =
\frac{G \rho_0}{3 \tilde{H}^2}
\dot\omega_0\frac{\frac{3}{4}\dot\omega_0 + \nu}{\left(
\frac{\dot\omega_0}{2} + \nu
\right)^2}
\label{Omega-11}
\end{equation}
These functions dependent on time as follow:
\begin{equation}
\delta_{c}(\mu,\tau) = \frac{3 \tilde{H}^2}{8 \pi G \rho_0} h^2(\mu,\tau).
\label{ro-cr-t}
\end{equation}
\begin{eqnarray}
\Omega =
\frac{16 G \rho_0}{3 \tilde{H}^2}
\dot\omega_0\left(\frac{3}{4}\dot\omega_0 + \nu\right) \, \cdot
\nonumber\\ \\
\frac{
      \left(1 \pm \frac{3}{4}\dot\omega_0\tau\right)
      \left[1 \pm \left(
      \frac{3}{4}\dot\omega_0+\nu
      \right) \tau \right]
     }
     {
     \left\{
            \pm 4 \left(
                        1 \pm \frac{3}{4}\dot\omega_0\tau
                   \right)
            \left(
                  \frac{3}{4}\dot\omega_0 + \nu
            \right)
     \mp \dot\omega_0
     \left[
           1 \pm
           \left(
                 \frac{3}{4}\dot\omega_0+\nu
           \right) \tau
     \right]
     \right\}^2
     }
\label{omega-3}
\end{eqnarray}
%


\section{FRW model} \label{nabl}

Every nonhomogeneous solution of the LTB model must include the FRW model
as a limited case
when the density and Habble function are
not dependent on space coordinate for
all moments of time.
Let's study in which case the present solution is
reduced to the FRW model?
Only one condition satisfied this request:
\begin{equation}
\omega_0 = const \qquad \mbox{so,} \quad \nu_0(\mu) = 0.
\label{vu = 0}
\end{equation}

It goes from (\ref{nu-def}) and (\ref{vu = 0}) that in this case
\begin{equation}
8 \pi \delta = \frac{3}{4} \left(
\frac{\dot\omega_0}{1 \pm \frac{3}{4}\dot\omega_0\tau}
\right)^2,
\label{rho_1_2}
\end{equation}
\begin{equation}
h = \pm\frac{1}{2} \frac{\dot\omega_0}{1 \pm\frac{3}{4}\dot\omega_0\tau}.
\label{Habbl_1_2}
\end{equation}
>From these equations it goes:
\begin{equation}
8 \pi \delta = 3 h^2,
\label{Habbl_1_3}
\end{equation}
We obtain the meaning of the $\omega_0$ from (\ref{rho_1_2}):
\begin{equation}
\dot\omega_0 = \pm 2 \sqrt{\frac{8 \pi}{3}}.
\label{chislo}
\end{equation}
$\nu = 0$ together with (\ref{chislo}) give
\begin{equation}
\Omega = \frac{G \rho_0}{\tilde{H}^2}.
\label{H-H}
\end{equation}
%
%

%

\section{Acknowledgements}

I'm grateful to Prof. Arthur D. Chernin for encouragement and discussion.
Dr. Yurij Barishev has initiated my interest to the modern Cosmology as
fractal structure of the Universe and interpretation of the observations by
calculation redshift in the LTB model.
This paper was financially supported by "COSMION" Ltd., Moscow.

\end{document}